\documentstyle[prl,aps,multicol,epsfig]{revtex}

\input epsf.tex

\newcommand{\be}{\begin{equation}}
\newcommand{\ee}{\end{equation}}
\newcommand{\ba}{\begin{eqnarray}}
\newcommand{\ea}{\end{eqnarray}}
\newcommand{\ban}{\begin{eqnarray*}}
\newcommand{\ean}{\end{eqnarray*}}
\newcommand{\braket}[2]{\mbox{$ \langle #1 | #2 \rangle $}}

\newcommand{\ket}[1]{\mbox{$ \left| #1 \right\rangle $}}

\newcommand{\demi}{\frac{1}{2}}

\newcommand{\one}{\leavevmode\hbox{\small1\normalsize\kern-.33em1}}

\begin{document}

\title{Towards practical and fast Quantum Cryptography}
\author{Nicolas Gisin$^1$, Gr\'egoire Ribordy$^2$, Hugo Zbinden$^1$,
Damien Stucki$^1$, Nicolas Brunner$^1$, Valerio Scarani$^1$}
\address{
$^1$ Group of Applied Physics, University of Geneva, 20, rue de
l'Ecole-de-M\'edecine, CH-1211 Geneva 4, Switzerland\\
$^2$ idQuantique, 3, Chemin de la Marbrerie, CH-1227 Carouge,
Switzerland}
\date{\today}
\maketitle \maketitle

\begin{abstract}

We present a new protocol for practical quantum cryptography,
tailored for an implementation with weak coherent pulses. The key
is obtained by a very simple time-of-arrival measurement on the
{\em data line}; an interferometer is built on an additional {\em
monitoring line}, allowing to monitor the presence of a spy (who
would break coherence by her intervention). Against zero-error
attacks (the analog of photon-number-splitting attacks), this
protocol performs as well as standard protocols with strong
reference pulses: the key rate decreases only as the transmission
$t$ of the quantum channel. We present also two attacks that
introduce errors on the monitoring line: the intercept-resend, and
a coherent attack on two subsequent pulses. Finally, we sketch
several possible variations of this protocol.
\end{abstract}

\begin{multicols}{2}

\section{Introduction}

Quantum cryptography \cite{review}, or quantum key distribution
(QKD), is probably the most mature field in quantum information,
both in theoretical and in experimental advances. On the
theoretical side, almost all QKD protocols have been proven to
provide unconditional security in some regime; on the practical
side, QKD has already reached the stage of commercial prototypes.
Still, much work is needed. A big task consists in bringing both
theory and applications in contact again: practical QKD systems do
not fulfill all the requirements of unconditional security proofs
(or, if you prefer, these proofs are still too abstract to cope
with a practical system). Here, we address a different question:
we aim for the {\em most practical} QKD system. Instead of looking
for a new implementations of known protocols, we choose to start
from scratch by inventing a new protocol. There are two basic
requirements:
\begin{itemize}

\item The protocol must be easily implementable, say with the
smallest number of standard telecom devices. Note that this
requirement, as a side benefit, may simplify security studies: we
have learnt in the recent years that any optical component can be
regarded as a "Trojan horse" because of its imperfections
\cite{makarov}.

\item The security of the system must be guaranteed by quantum
physics, thence in some way quantum coherence must play a role.

\end{itemize}

The goal of this paper is to illustrate this program by presenting
such a system. the key is created in a {\em data line} that is
probably the simplest one can think of --- just measure the time
of arrival of weak pulses. The intervention of a spy is checked
interferometrically in a {\em monitoring line}. In Section
\ref{proto}, we define precisely the protocol and stress its
advantages. In Section \ref{attacks}, we present a quantitative
study of security. Finally, Section \ref{secvar} presents a number
of possible variations on the main idea.

\section{The protocol}
\label{proto}

\begin{center}
\begin{figure}
\includegraphics[width=8cm]{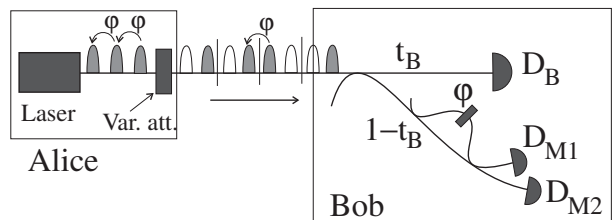} \caption{Scheme of the protocol.
Bob reads the raw key in detector $D_B$, the monitoring line
checks for the breaking of quantum coherence due to an
eavesdropper. See text for details} \label{figproto}
\end{figure}
\end{center}

\subsection{The source}

Alice uses a mode-locked laser, producing pulses of mean
photon-number $\mu$ that are separated by a fixed and well-defined
time $\tau$; with a variable attenuator, she can blocks some of
the pulses (note that a more economical source would just consist
of a cw laser followed by the variable attenuator). Each logical
bit is encoded in a two-pulse sequence according to the following
rules:
\ba \ket{0_A}&=&\ket{\sqrt{\mu}\,e^{i(2k-1)\varphi}}_{2k-1}\ket{0}_{2k}\,,\\
\ket{1_A}&= &\ket{0}_{2k-1}\ket{\sqrt{\mu}\,
e^{i\,(2k)\varphi}}_{2k}\,.\ea For instance, the eight-pulse
sequence drawn in Fig. \ref{figproto} codes for the four-bit
string 0100 (read in temporal order, that is, from right to left).
For small $\mu$, the states $\ket{0_A}$ and $\ket{1_A}$ have a
large overlap because of their vacuum component. Since the laser
is mode-locked, there is a phase coherence between any two
non-empty pulses. Leaving a more general discussion for Section
\ref{secvar}, we focus on the case where bit number $k$ is 1 and
bit number $k+1$ is 0, like bits number 2 and 3 of \ref{figproto}.
Then {\em across the bit-separation} there is a phase coherence:
\ba \ket{\sqrt{\mu}\,e^{i(2k)\varphi}}_{2k}\,
\ket{\sqrt{\mu}\,e^{i(2k+1)\varphi}}_{2k+1}\,. \ea Note that the
choice of the value of $\varphi$ is arbitrary, so from now on we
set $\varphi=0$.

\subsection{The "data line"}

The pulses now propagate to Bob, on a quantum channel
characterized by a transmission $t=10^{-\alpha\,d/10}$ (a typical
value for $\alpha$ in optical fibers is 0.2 dB/km). Bob's setup
first splits the pulses using a non-equilibrated beam-splitter
with transmission coefficient $t_B$. The pulses that are
transmitted are used to establish the raw key ({\em data line}).
To obtain the bit value, Bob has to distinguish unambiguously
between the two non-orthogonal states \ba
\ket{0_B}&=&\ket{\alpha}_{2k-1}\ket{0}_{2k}\,,\\
\ket{1_B}&= &\ket{0}_{2k-1}\ket{\alpha}_{2k}\ea with $\alpha =
\sqrt{\mu\,t\,t_B}$ --- we have omitted the phase due to the free
propagation, which is the same for all pulses. As well-known,
unambiguous discrimination between two pure states can succeed
with probability $p_{ok}=1-|\braket{0_B}{1_B}|$; in the present
case, the overlap is $|\braket{0_B}{1_B}|=e^{-|\alpha|^2}$, and
consequently $p_{ok}=1-e^{-\mu \,t\,t_B}$. Now, there is an
obvious way to achieve this result: photon counting with a perfect
detector, because $p_{ok}$ is just the probability that the
detector will detect something. The realistic situation where the
detector has a finite efficiency $\eta$ can be modelled by an
additional beam-splitter with transmittivity $\eta$ followed by a
perfect detector; in this case, $\eta$ appears in the exponent as
well. In conclusion, the optimal unambiguous discrimination
between $\ket{0_B}$ and $\ket{1_B}$ is achieved by the most
elementary strategy, simply try to detect where the photons are.
Later, Bob must announce Alice which items he has detected: this
is how Alice and Bob establish their raw key. Note that no error
is expected on this line, if the switch is perfect and in the
absence of dark counts of the detector: a bit-flip is impossible
because it would correspond to a photon jumping from a time-bin to
another.

Note that the simplicity of Bob's data line has concrete practical
advantages. There are no lossy and active elements. Hence, the
transmission range can be increased and no random number generator
is needed.

As for the data line, our protocol is similar to the one of
Debuisschert and Boucher \cite{thales}. However there, the
security was obtained by the overlap in time between the pulses
coding for different bits. Here, we use rather the monitoring line
described in the next paragraph.

\subsection{The "monitoring line"}

The pulses that are reflected at Bob's beam-splitter go to an
interferometer that is used for monitoring Eve's presence ({\em
monitoring line}). Here is where quantum coherence plays a role.
Let $\alpha_j$ be the amplitude of pulse $j$ entering the
interferometer: in particular, $|\alpha_j|^2$ is either 0 or
$\mu\,t\,(1-t_B)$; and if both $\alpha_j$ and $\alpha_{j+1}$ are
non-zero, then $\alpha_{j+1}=\alpha_j$. After the interferometer,
the pulses that reach the detectors at time $j+1$ is given by \ba
\ket{D_{M1}}&=& \ket{i\frac{\alpha_j +
\alpha_{j+1}}{2}}\\
\ket{D_{M2}}&=& \ket{\frac{-\alpha_j + \alpha_{j+1}}{2}}\,.\ea
Now, if either $\alpha_j$ or $\alpha_{j+1}$ are zero, then
$|D_{M1}|^2 = |D_{M2}|^2 = \demi \mu\,t\,(1-t_B)$; i.e.,
conditioned to the fact that a photon takes the monitoring line,
the probabilities of detecting it in either detector is $\demi$.
However, if both $\alpha_j$ and $\alpha_{j+1}$ are non-zero, then
$|D_{M1}|^2 = \mu\,t\,(1-t_B)$ and $|D_{M2}|^2 = 0$: only detector
$D_{M1}$ can fire. Consider then again the case where bit number
$k$ is 1 and bit number $k+1$ is 0: as we said above, in this case
the two consecutive pulses $2k$ and $2k+1$ are non-empty. This
means that, {\em if coherence is not broken}, detector $D_{M2}$
cannot fire at time $2k+1$. If Eve happens to break the coherence
by reading the channel, it could be detected this way.

Actually, it turns out that, as just described, the protocol is
insecure: Eve can make a {\em coherent} measurement of the number
of photons in the two pulses across the bit-separation. With such
an attack, she would not break the coherence, thus introduce no
errors in the monitoring line, and obtain almost full information
(see next Section for more details). There are several ways of
countering this attack: here, we make use of {\em decoy
sequences}, inspired by the idea of "decoy states" introduced by
Hwang \cite{hwang} and by Lo and co-workers \cite{lo}, but
different in its implementation. The principle is the following:
with probability $f$, Alice leaves both the $(2k-1)$-th and the
$2k$-th pulses non-empty. A decoy sequence does not encode a bit
value (in contrast to the decoy states of \cite{hwang,lo} that
still encode a state, but in a different way):  thence, if the
item is detected in the data line, it will be discarded in public
discussion. However, if a detection takes place in the monitoring
line at time $2k$, then it must be in detector $D_{1M}$ because of
coherence. Now Eve can no longer pass unnoticed: if she attacks
coherently across the bit separation, then she breaks the
coherence of the decoy sequences; if she attacks coherently within
each bit, then she breaks the coherence across the separation;
finally, if she makes a coherent attack on a larger number of
pulses, then she breaks the coherence in fewer positions but gets
much less information.

Thus, errors are rare: they appear only in the monitoring line,
and just for a fraction of the whole cases. Still, one can
estimate the error (thence, the coherence of the channel) in a
reasonable time, if the bit rate is high.

\subsection{Summary of the protocol}

Let's summarize the protocol before moving to a more quantitative
study of security:
\begin{enumerate}
\item Alice prepares "bit 0" with probability $\frac{1-f}{2}$,
"bit 1" with probability $\frac{1-f}{2}$ and the decoy sequence
with probability $f$. This is repeated a large number of times.

\item At the end of the exchange, Alice reveals the items
$\{k_d\}$ corresponding to a decoy sequence. Bob removes all the
detections at times $2k_d-1$ and $2k_d$ from his raw key, and he
looks whether detector $D_{2M}$ has ever fired at times $2k_d$.
This way, Alice and Bob estimate the break of coherence of decoy
pulses.

\item On the remaining fraction of sent bits $1-f$, Bob reveals
the times $2k+1$ in which he had a detection in $D_{2M}$. Alice
verifies if some of these items correspond to a bit sequence
"1,0"; thus, Alice and Bob estimate the break of coherence across
the bit-separation.

\item Finally, Bob reveals the items that he has detected in the
data line. Alice and Bob run error correction and privacy
amplification on these bits and end up with a secret key.

\end{enumerate}

Should one say in one sentence where the improvement lies, here it
is: {\em one can define a very simple data line and protect it
quantum-mechanically}.

At this point, two important remarks can be done. First, this
protocol cannot be analyzed in terms of qubits. This is obvious,
because any bits and coherence are checked on differently defined
pairs. In particular, there is not a "natural" single-photon
version of the protocol (simply replace non-empty coherent state
with one-photon Fock states would be dramatic, since all the
sequences would become orthogonal). The second remark is the
answer to a possible question. With the idea of a simple data line
for key creation, and a "complementary" line for monitoring, one
may implement a version of the BB84 protocol: Alice and Bob agree
to produce the key using only the $Z$ basis; sometimes Alice
prepares one of the eigenstates of the $X$ basis that acts as a
decoy state. Which are the advantages of our protocol? We are
going to see that our protocol is much more robust against attacks
at zero errors (the analog of photon-number-splitting attacks).

\section{Quantitative analysis of security}
\label{attacks}

For a reasonable comparison with experiment, we must introduce the
following parameters
\begin{itemize}

\item The visibility $V$ of the monitoring interferometer, whence
the probability that $D_{2M}$ fires in a time corresponding to a
coherence is $\frac{1-V}{2}$ instead of zero. We suppose that Eve
can take advantage of these imperfections: for instance, if the
reduced visibility is due to $\varphi\neq 0$ in the
interferometer, Eve can systematically correct for this error by
displacing the pulses, and then reproduce $V$ by adding errors in
a way that is profitable for her.

\item The imperfections of the three Bob's detectors, supposed to
be identical for simplicity: the quantum efficiency $\eta$ and the
probability per gate of a dark count $p_d$. Typical values are
$\eta=10\%$, $p_d=10^{-5}$. These imperfections are not given to
Eve (see Section \ref{secvar} on the possibility that Eve forces a
detection, thus effectively setting $\eta=1$ for some pulses).

\end{itemize}

For simplicity in writing, we make all the quantitative analysis
in the limit of small mean photon-number in Bob's channel, that is
$\mu t<<1$.

\subsection{Parameters Alice-Bob on the data line}
\label{ssab}

First, we compute the parameters of Alice-Bob on the data line.
Bob's detection rate in $D_B$, once decoy sequences are removed,
is \ba R_B&=& \big[\mu T \,+\,(1-\mu T)p_d\big]\,(1-f) \ea where
$T=t\,t_B\,\eta$. In other words, $R_B$ times the number of
two-pulse sequences sent by Alice is the length of the raw key.

If we assume that the switch prepares really empty pulses when it
is closed, the error expected in this line is only due to the dark
counts of the detectors: \ba Q&=& \frac{\demi\,(1-\mu
T)p_d(1-f)}{R_B} \ea because dark counts may make the detector
fire at both times with equal probability. The mutual information
Alice-Bob in bits per sent photon is thence \ba
I(A:B)&=&R_B\,\big[1-H(Q)\big]\,. \label{iab}\ea In what follows,
we shall concentrate on attacks by Eve that do not modify $Q$.
Before moving to that, let's have a look at the monitoring line as
well.

\subsection{About the monitoring line}

In the presence of dark counts and reduced visibility, the
meaningful detection probabilities in $D_{M1}$ and $D_{M2}$,
neglecting double counts are the following \cite{note1}: \ba
\mbox{Time $2k$,
decoy seq. }&:& \;R^{d}_{1,2}\,=\,R^{d}_{M1,2}\,f \label{rd}\\
\mbox{Time $2k+1$, seq. "1,0" }&:&
\;R^{10}_{1,2}\,=\,R^{10}_{M1,2}\,\frac{1-f}{4}\label{r10}\ea
where, denoting $\tilde{T}=t\,(1-t_B)\,\eta$, we have defined \ba
R^{\,x}_{M1,2}&=&\mu\tilde{T}\frac{1\pm
V_x}{2}+\left(1-\mu\tilde{T}\frac{1\pm V_x}{2}\right)p_d\,. \ea
Contrary to the errors due to dark counts, the departure from
perfect visibility will be entirely attributed to Eve. This is why
we consider {\em a priori} different values $V_d$ and $V_{10}$ for
the visibility in the two cases: as we shall see, Eve's attacks
may be different.

\subsection{Eve's attacks}

If Bob's detector has dark counts, $I(B:E)$ is smaller than
$I(A:E)$ for a prepare-and-measure scheme, because even if Eve
knows perfectly what Alice has sent, she cannot know whether Bob
has detected a photon or has had a dark count. Thus in our case,
the Csiszar-K\"orner bound \cite{csi78} that gives an estimate of
the extractable secret key rate becomes \ba\label{ckbound}
R&\geq&I(A:B)\,-\,\min \big\{I(A:E),I(B:E)\big\}\nonumber\\ &=&
I(A:B)\,-\,I(B:E)\,. \ea Therefore, we have to compute the mutual
information Bob-Eve.

\begin{center}
\begin{figure}
\includegraphics[width=8cm]{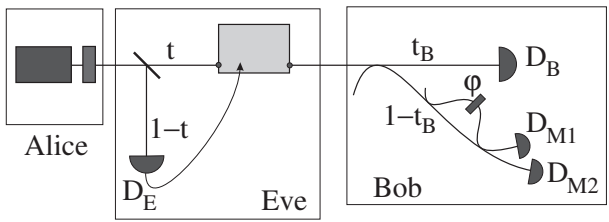} \caption{Scheme with Eve.}
\label{figeve}
\end{figure}
\end{center}

The kind of attacks by Eve that we consider is sketched in Fig.
\ref{figeve}. We can give Eve all the losses in the line, that is,
we can suppose that Eve removes a fraction $1-t$ of the photons,
and forwards the remaining fraction $t$ to Bob on a lossless line.
We are going to study:
\begin{itemize}
\item An attack in which Eve can gain information without
introducing errors. This is related to the losses on the line; it
is the analog of the usual photon-number-splitting attack
\cite{pns,pns2}, but is a different attack and less powerful.

\item Eve can immediately know if the previous attack was
successful or not; in the case it wasn't, we consider further the
possibility of attacks that introduce errors in the monitoring
line (but still no errors in the data line). Specifically, we
study a usual intercept-resend strategy, and a more clever attack
which is performed coherently on two subsequent pulses across the
bit-separation.
\end{itemize}

\subsubsection{Eve's attack without errors}

In the case of BB84 and many other protocols, Eve can exploit
multi-photon pulses in a lossy line to perform the {\em
photon-number-splitting attack} \cite{pns,pns2}: she counts the
photons in each pulse, and whenever this number is larger than
one, she keeps one photon in a quantum memory and forwards the
remaining photons to Bob on a lossless line. As such, this attack
is not error-free in the present protocol: counting the photons in
each pulse breaks the coherence between successive pulses, thus
introducing errors in the monitoring line --- actually, because of
the peculiar encoding of the bits, this attack reduces here to the
intercept-resend, see below.

More subtle is the analysis of a practical version of the attack
using {\em cascaded beam-splitters} \cite{felix}: Eve uses a
highly unbalanced BS, with transmission $1-\varepsilon$ and
reflection $\varepsilon$; if she has a detection, she forwards the
remaining photons to Bob; otherwise, she begins anew, and so on
until the losses that she introduces reach the transmission $t$ of
the quantum channel. The advantage of this strategy is that, in
the presence of two or more photons, it is very rare that more
than one photon is coupled into Eve's detector. Indeed, this
beam-splitting attack approximates a photon-counting. In our case,
this strategy will introduce errors in the monitoring line as
well: it does not modify the relative phase, but the relative
intensity between subsequent pulses, thus leading to an
unbalancing of the interferometer. The full analysis of such a
strategy will be studied in a further work.

In summary, both the ideal photon-number-splitting and its
approximate implementation through cascaded beam-splitters do not
rank among the zero-error attacks against our protocol. In fact,
Eve can only perform the basic {\em beam-splitting attack}: she
removes a fraction $1-t$ of the photons, and transmits the
remaining fraction $t$ to Bob on a lossless line. With the
fraction that she has kept, the best thing Eve can do is just to
measure them (recall the argument about optimal unambiguous state
discrimination). This way, she detects $\mu(1-t)$ photons per
pulse. When Eve has a detection in $D_E$, she knows the bit that
Alice has sent. Then she lets the remaining part of the pulse
travel to Bob on the lossless line (the grey box of Fig.
\ref{figeve} is simply a line). Bob detects something exactly as
if Eve had not been there. So \ba I(B:E|D_E)&=&I(A:B)\,.
\label{iede}\ea In summary, Eve knows a fraction $\mu(1-t)$ of the
key just because of the losses in the quantum channel: this
fraction must always be subtracted in privacy amplification.

It is instructive to compute the optimal value of $\mu$ under the
assumption that Eve introduces no errors, and neglecting dark
counts ($Q=0$). In BB84, this value is $\mu_{BB84}=t$, giving
$R_{BB84}=\frac{1}{4}\eta\, t^2$ \cite{mubb}. Here, Alice and Bob
must maximize $R$ given in (\ref{ckbound}); using (\ref{iab}) and
(\ref{iede}), this reads \ba R&=&
\mu\,t\,t_B\,\eta\,(1-f)\,\big(1-\mu(1-t)\big)\,. \ea The
optimization $dR/d\mu=0$ is readily done and yields \ba
\mu_{opt}&=& \frac{1}{2(1-t)}\ea whence \ba
R_{opt}&=&\frac{1}{4(1-t)}\,t\,t_B\,\eta\,(1-f)\,.\ea This is an
important improvement over BB84: $\mu_{opt}$ is large and is
basically constant with decreasing $t$ (long quantum channels); as
a consequence, the secret-key rate decreases only {\em linearly}
(and not quadratically) with $t$. This is the same improvement
that can be obtained by using decoy states \cite{lo} or a strong
reference pulse \cite{koashi}; note however that the hardware is
much simpler here.

\begin{center}
\begin{figure}
\includegraphics[width=8cm]{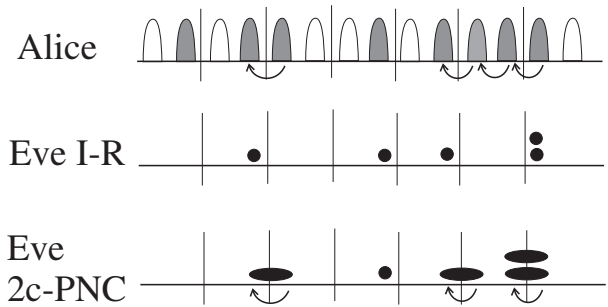} \caption{Comparison of two
attacks that introduce errors. In the I-R attack, Eve prepares a
sequence of localized Fock states, thus breaking the coherence
everywhere. In the 2c-PNC attack, Eve prepares a sequence of Fock
states that are delocalized across the separation of bits: only
the coherence of decoy sequences is broken. Note that arrows
denote only coherence between subsequent pulses, the one checked
by the interferometer; however, on the original sequence, all the
non-empty pulses are coherent with one another, while in the
sequences after Eve's attack only the indicated coherence
remains.} \label{figattacks}
\end{figure}
\end{center}

When Eve's detector $D_E$ does not fire, which happens with
probability $1-\mu(1-t)$, Eve must perform some attack on the
pulses flying to Bob if she wants to gain some information. These
attacks will certainly introduce errors, either in the data line
or in the monitoring line. In the following, we present two such
attacks (Fig. \ref{figattacks}): a basic intercept-resend (I-R),
and a photon-number-counting attack performed coherently on two
subsequent pulses across the bit-separation (2c-PNC).

\subsubsection{Eve's attack with errors (I): intercept-resend}

Let's begin with the intercept-resend (I-R) strategy. Eve simply
detects the pulse flying to Bob: her detector will fire with
probability $\mu t$, and in this case she prepares a {\em
single-photon} in the good time-bin and forwards it to Bob.
Obviously, both $R_B$ and $Q$ are unchanged under this strategy.

Note that $R_B$ will be the sum of three terms: Eve has detected
and Bob detects too; Eve has detected and Bob has a dark count;
Eve has not detected and Bob has a dark count. Now, Eve can
distinguish the last one from the two first, and she knows that in
the last case she has no information on Alice's and Bob's bits. So
\ba I(B:E|IR)&=&[R_B-(1-\mu t)p_d(1-f)]\big[1-H(Q')\big] \ea where
\ba Q'&=&\frac{\demi(1-t_B\eta)p_d}{t_B\eta + (1-t_B\eta)p_d}
\label{qprime}\ea 
is the fraction of Bob's detection on which she has the wrong
result.

Of course, the I-R attack breaks all the coherences (see Fig.
\ref{figattacks}), and will therefore introduce errors in the
monitoring line. Specifically, whenever Eve has performed the I-R
attack, \ba V_{d|IR}=V_{10|IR}&=&0\,.\ea

\subsubsection{Eve's attack with errors (II): 2-coherent PNC}

In the absence of decoy sequences, Eve may obtain information
without introducing errors in the monitoring line, by counting the
number of photons coherently between two pulses, not within each
bit but across the separation line (see Fig. \ref{figattacks}).
This attack does not break the phase between these pulses. Of
course, if Eve finds $n>0$ photons, on the spot she does not know
to which bit the photon belongs; but she will learn it later, by
listening to Bob's list of accepted bits \cite{note0}. Actually,
in some very rare case Eve still does not get any information: if
Eve prepares $n>0$ photons in two successive two-pulse sequences,
and Bob accepts a detection in the bit common to both sequences,
Eve has no idea of his result. However, since such cases are rare,
we make the conservative assumption that Eve always gets full
information. Thus \ba I(B:E|2cPNC)&=&I(B:E|IR)\,. \ea Eve has
introduced errors in the monitoring line {\em only} in the items
corresponding to decoy sequences: \ba
V_{d|2c}=0&\,,\;&V_{10|2c}=1\,.\ea

\subsubsection{Collecting everything}

We can now collect all that we know form this analysis of
security. The parameters describing Alice and Bob are those listed
in paragraph \ref{ssab} above. Let $p_{IR}$ and $p_{2c}$ be the
probabilities that Eve performs the I-R attack, resp. the 2c-PNC
attack. Recall that she performs these attacks only when her
detector $D_E$ did not fire, so $p_{IR}+p_{2c}\leq 1-\mu(1-t)$.
Then: \ba I(B:E)&=&\mu(1-t)\,R_B\, \big[1-H(Q)\big]\,+\nonumber\\
&+& (p_{IR}+p_{2c})\,R_B'\,\big[1-H(Q')\big] \ea with the notation
$R_B'=R_B-(1-\mu t)p_d(1-f)$ and with $Q'$ given in
(\ref{qprime}). This is the expression that must be inserted into
(\ref{ckbound}) to obtain the extractable secret key rate (in
other terms, this is the quantity to be corrected by privacy
amplification).

On the monitoring line, Alice and Bob measure (\ref{rd}) and
(\ref{r10}), whence they extract $V_d$
and $V_{10}$ that inform them about Eve's attacks according to
\ba V_d &=& 1-p_{IR}-p_{2c}\,, \\
V_{10} &=& 1-p_{IR}\,.\ea Note in particular that, if Alice and
Bob find $V_d=V_{10}$, they can conclude that Eve has not used the
2c-PNC attack --- by the way, this is why we presented the
analysis of the I-R strategy, obviously worse than 2c-PNC from
Eve's standpoint: in a practical experiment, $V_d=V_{10}$ is very
likely to hold (after all, Eve is not there...). Therefore,
formulae for the I-R attack will be useful in the analysis of
experimental data.

\section{Variations and open questions}
\label{secvar}

\subsection{Variations}

Here are a few ideas of variations in the protocol, that may have
some additional benefit and require further study:

\begin{itemize}

\item Alice may change during the protocol the definition of the
pulses that define a bit. If there is a convenient fraction of
decoy sequences, Eve has no way of distinguishing a priori which
pairs of successive pulses encode a bit. This way, the effect of
the 2c-PNC attack becomes equally distributed among decoy and
"1,0" sequences, i.e. $V_d=V_{10}= 1-p_{IR}-\demi p_{2c}$.
Moreover, whenever Eve attacks the bit instead of attacking across
the bit-separation, she cannot gain any information.

\item In fact, nothing forces to define bits by subsequent pulses:
Alice and Bob can decide later, adding a sort of "sifting" phase
to the protocol. This means that Alice can now send whatever pulse
sequence, she is no longer restricted to those that define a bit
or a decoy sequence. It is not clear whether this modification
helps, if the hardware is kept unchanged: Alice and Bob still
check only the coherence between subsequent pulses; moreover,
sifting means additional losses and additional information
revealed publicly.

\item Instead of introducing decoy sequences as we did above, one
may study the effect of decoy pulses with different intensities,
as proposed by Hwang and by Lo and co-workers to protect the BB84
protocol against the photon-number-splitting attack
\cite{hwang,lo}.

\end{itemize}

\subsection{Possible loopholes}

The difficulty in assessing the security of practical QKD, is the
huge number of imperfections that may hide loopholes for security.
These imperfections exist in all implementations and for all
protocols, but their effect and the corresponding protection may
vary. Here we present some of these.

\begin{itemize}

\item About Trojan-horse and similar realistic attacks
\cite{makarov}: Alice's setup must be protected against
Trojan-horse attacks, with the suitable filters, isolators etc. In
Bob's setup nothing is variable; however, one must prevent the
possible light emission from avalanche photodiodes to become
available to Eve: if the firing of a detector can be seen from
outside Bob, the protocol becomes immediately insecure.

\item After a detection, Bob's detectors are blind during some
time. In particular, if Eve happens to know when Bob's detector
$D_{M2}$ has fired, she can attack strongly the subsequent pulses
because no error will be detected then, and gain one bit (just
one, because when $D_B$ has fired, then it has a dead time as
well). For the security of the protocol, Eve must have no way of
assessing the firing of a detector, and Bob must announce publicly
this information only after the detector is ready again. This may
imply some suitable synchronization in the software, or more
simply, to shut $D_B$ as long as $D_{M2}$ has not recovered. The
nuisance depends of course on the ratio between the raw bit rate
and the dead time.

\item In all this paper, we have considered only the case where
Eve does not change Bob's detection rates in the data line and in
the monitoring line. By sending out stronger pulses, Eve might
force the detection of those items on which she has full
information; but in turn, she would increase the rate of double
counts among Bob's detectors. This effect must be quantified, and
the number of double, or even triple, counts must be monitored
during the experiment.

\end{itemize}

\section{Conclusion}

In conclusion, we have presented a new protocol for quantum
cryptography whose realization is much simpler than that of
previously described ones. Specifically, Bob's station is such
that losses are minimized and no dynamical component is needed.

We thank H.-K. Lo for stimulating comments. We acknowledge
financial support from idQuantique and from the Swiss NCCR
"Quantum photonics".

\end{multicols}


\begin{thebibliography}{10}
\bibitem{note1} Just to verify normalization, we notice that the rate of
uninteresting detections are: for detection at time $2k+1$
corresponding to both sequences "0,0" and "1,1",
$\left[\frac{\mu}{2}\tilde{T}\demi
+\left(1-\frac{\mu}{2}\tilde{T}\demi\right)p_d\right]\frac{1-f}{4}$
in each detector; for detection at time $2k+1$ corresponding to
sequence "0,1", $p_d\frac{1-f}{4}$ in each detector. Thus, the sum
of all detections of non-decoy sequences sums up to
$[\frac{\mu}{2}\tilde{T} +(2-\frac{\mu}{2}\tilde{T})p_d](1-f)$ as
it should, because in average there are $\frac{\mu}{2}\tilde{T}$
photons at each time in non-decoy sequences and there are two
detectors in which a dark count may happen.
\bibitem{mubb} A. Niederberger, V. Scarani, N. Gisin,
quant-ph/0408122. Recall that we have assumed that the errors in
the detectors are not given to Eve. In the most conservative case
where all the errors are given to Eve, the estimates of $\mu$ must
be multiplied by $\eta$ \cite{pns2}.
\bibitem{note0} For simplicity, we consider the case where Eve
applies this attack on several consecutive items, see Fig.
\ref{figattacks}. One might as well consider the case where Eve
applies the attack, sometimes, on just a single item. In this
case, she learns full information (apart from dark counts) if she
detects 0 photons and Bob accepts one of the two bits; while she
does not learn anything if she has detected some photons and Bob
accepts. The full analysis is longer than the one presented, and
leads to the same yield in terms of information.
\end{thebibliography}
\end{document}